\documentclass[12pt,english,american]{article}
\usepackage[T1]{fontenc}
\usepackage[utf8]{inputenc}
\usepackage[skip=\smallskipamount]{parskip}
\usepackage{graphicx}
\usepackage{setspace}
\makeatletter

\providecommand{\tabularnewline}{\\}

\makeatother

\usepackage{babel}
\begin{document}
\title{An Improved Satterthwaite Effective Degrees of Freedom Estimator for
Weighted Syntheses of Variance}
\author{Matthias von Davier\thanks{Monan Professor in Education, Executive Director, TIMSS and PIRLS
International Study Center at Boston College. email: vondavim@bc.edu}}
\date{May 15, 2025}
\maketitle
\begin{abstract}
This article presents an improved approximation for the effective
degrees of freedom in the Satterthwaite (1941, 1946) method which
estimates the distribution of a weighted combination of variance components
The standard Satterthwaite approximation assumes a scaled chisquare
distribution for the composite variance estimator but is known to
be biased downward when component degrees of freedom are small. Building
on recent work by von Davier (2025), we propose an adjusted estimator
that corrects this bias by modifying both the numerator and denominator
of the traditional formula. The new approximation incorporates a weighted
average of component degrees of freedom and a scaling factor that
ensures consistency as the number of components or their degrees of
freedom increases. We demonstrate the utility of this adjustment in
practical settings, including Rubin's (1987) total variance estimation
in multiple imputations, where weighted variance combinations are
common. The proposed estimator generalizes and further improves von
Davier's (2025) unweighted case and more accurately approximates synthetic
variance estimators with arbitrary weights.

\end{abstract}

\section{Introduction:}

This article presents an improved approximation for the effective
degrees of freedom in the Satterthwaite (1941, 1946) method, which
is used for estimating the distribution of a weighted combination
of variance components. The standard Satterthwaite approximation assumes
a scaled chi-square distribution for the composite variance estimator
but is known to be biased downward when component degrees of freedom
are small. Building on recent work by von Davier (2025), we propose
an adjusted estimator that corrects for this bias by modifying both
the numerator and denominator of the traditional formula. The new
approximation incorporates a weighted average of component degrees
of freedom and a scaling factor that ensures consistency as the number
of components or their degrees of freedom increase. We demonstrate
the utility of this adjustment in practical settings, including Rubin’s
(1987) total variance estimation in multiple imputation, where weighted
variance combinations are common. The proposed estimator generalizes
von Davier’s (2025) unweighted case and provides a more accurate approximation
for synthetic variance estimators with arbitrary weights. 

The Satterthwaite effective degrees of freedom approximation is defined
for a synthesis of variance estimators. More specifically, for an
expression
\[
\sigma_{*}^{2}=\sum_{k=1}^{K}w_{k}\sigma_{k}^{2}
\]

Satterthwaite (1941, 1946) argues that an approximate $\chi^{2}$
distribution can be assumed for variance estimators that are defined
as a linear combination of variances, and proposes the following approximation
of the degrees of freedom 
\[
\nu_{*}\approx\frac{\left[\sum_{k}w_{k}S_{k}^{2}\right]^{2}}{\sum_{k}w_{k}^{2}\frac{\left(S_{k}^{2}\right)^{2}}{\nu_{k}}}
\]

with components 
\[
\nu_{k}\frac{S_{k}^{2}}{\sigma_{k}^{2}}\sim\chi_{\nu_{k}}^{2}
\]
that are used to calculate
\[
S_{*}^{2}=\sum_{k=1}^{K}S_{k}^{2}.
\]

The basic tenet of Satterthwaite (1941,1946) is that we may assume
that 
\[
\nu_{*}\frac{S_{*}^{2}}{\sigma_{*}^{2}}\sim\chi_{\nu_{*}}^{2}.
\]

\subsection{Things to Note}

For $\chi^{2}-$distributed random variables it is well known that
\[
E\left[\nu_{*}\frac{S_{*}^{2}}{\sigma_{*}^{2}}\right]=\nu_{*}
\]
and 
\[
Var\left[\nu_{*}\frac{S_{*}^{2}}{\sigma_{*}^{2}}\right]=2\nu_{*}.
\]
We will use some well known identities to arrive at the main result.
First, recall that $Var\left(X\right)=E\left(X^{2}\right)-E\left(X\right)^{2}$
and that $Var\left(aX\right)=a^{2}Var\left(X\right)$ and that for
independent $X_{1},...,X_{K}$we have $Var\left(\sum_{k}w_{k}X_{k}\right)=\sum_{k}w_{i}^{2}Var\left(X_{k}\right).$

The following equivalencies hold
\[
\nu_{*}^{2}\frac{Var\left(S_{*}^{2}\right)}{\left(\sigma_{*}^{2}\right)^{2}}=2\nu_{*}\leftrightarrow\nu_{*}=2\frac{\left(\sigma_{*}^{2}\right)^{2}}{Var\left(S_{*}^{2}\right)}
\]

and 
\[
Var\left(S_{*}^{2}\right)=\sum_{k=1}^{K}w_{k}^{2}Var\left(S_{k}^{2}\right)=\sum_{k=1}^{K}w_{k}^{2}2\frac{\left(\sigma_{k}^{2}\right)^{2}}{\nu_{k}}
\]
since
\[
Var\left(S_{k}^{2}\right)=2\frac{\left(\sigma_{k}^{2}\right)^{2}}{\nu_{k}}
\]

which provides
\[
\nu_{*}=2\frac{\left(\sigma_{*}^{2}\right)^{2}}{\sum_{k=1}^{K}w_{k}^{2}2\frac{\left(\sigma_{k}^{2}\right)^{2}}{\nu_{k}}}=\frac{\left(\sum_{k=1}^{K}w_{k}\sigma_{k}^{2}\right)^{2}}{\sum_{k=1}^{K}w_{k}^{2}\frac{\left(\sigma_{k}^{2}\right)^{2}}{\nu_{k}}}.
\]

\section{Satterthwaite Effective d.f. and Closer Approximations}

Satterthwaite's (1946) estimator $\nu_{*,Satt}$ of the effective
degrees of freedom for composite variances $S_{*}^{2}$ was found
to be biased downwards when the components $S_{k}^{2}$ have small
or single degrees of freedom $\nu_{k}<<\infty$ or even $\nu_{k}=1$
(e.g., Johnson \& Rust, 1992; Qian, 1998). 

\subsection{Some Properties of the Weighted d.f. Estimator}

It is worth noting that the effect of the weights is invariant under
rescaling and it can be assumed that all weights are postive, $w_{k}>0$,
since zero weights would simply reduce the number of additive terms,
and negative weights could lead to a negative variance. It can be
shown easily that the same estimate will be obtained for two sets
of weights $w_{\cdot},w_{\cdot}^{\#}$ with $w_{k}^{\#}=cw_{k}$ for
all $k.$

\[
\frac{\left(\sum_{k=1}^{K}w_{k}^{\#}S_{k}^{2}\right)^{2}}{\sum_{k=1}^{K}\left(w_{k}^{\#}\right)^{2}\frac{\left(S_{k}^{2}\right)^{2}}{\nu_{k}}}=\frac{\left(\sum_{k=1}^{K}cw_{k}S_{k}^{2}\right)^{2}}{\sum_{k=1}^{K}c^{2}w_{k}^{2}\frac{\left(S_{k}^{2}\right)^{2}}{\nu_{k}}}=\frac{c^{2}\left(\sum_{k=1}^{K}w_{k}S_{k}^{2}\right)^{2}}{c^{2}\left[\sum_{k=1}^{K}w_{k}^{2}\frac{\left(S_{k}^{2}\right)^{2}}{\nu_{k}}\right]}=\frac{\left(\sum_{k=1}^{K}w_{k}S_{k}^{2}\right)^{2}}{\sum_{k=1}^{K}w_{k}^{2}\frac{\left(S_{k}^{2}\right)^{2}}{\nu_{k}}}
\]

So we may define 
\[
c=\min\left\{ w_{k}^{\#}:k=1,\dots,K\right\} 
\]
and can assume that we have at least one $w_{k}=1$, and $\forall k:w_{k}\ge1$
from now on. 

The same property can be shown for the $S_{k}^{2}$, use $r=\min\left\{ S_{k}^{2}:k=1,\dots,K\right\} $
and define 
\[
R_{k}^{2}=\frac{1}{r}S_{k}^{2}
\]
so that there exists a $k^{'}$with $R_{k^{'}}^{2}=1$ and $R_{l}^{2}>R_{k^{'}}^{2}$for
all $l\ne k.$ Note that 
\[
\frac{\left(\sum_{k=1}^{K}w_{k}R_{k}^{2}\right)^{2}}{\sum_{k=1}^{K}w_{k}^{2}\frac{\left(R_{k}^{2}\right)^{2}}{\nu_{k}}}=\frac{\left(\sum_{k=1}^{K}w_{k}\frac{1}{r}S_{k}^{2}\right)^{2}}{\sum_{k=1}^{K}w_{k}^{2}\frac{\left(\frac{1}{r}S_{k}^{2}\right)^{2}}{\nu_{k}}}=\frac{\frac{1}{r^{2}}\left(\sum_{k=1}^{K}w_{k}S_{k}^{2}\right)^{2}}{\frac{1}{r^{2}}\left[\sum_{k=1}^{K}w_{k}^{2}\frac{\left(S_{k}^{2}\right)^{2}}{\nu_{k}}\right]}=\frac{\left(\sum_{k=1}^{K}w_{k}S_{k}^{2}\right)^{2}}{\sum_{k=1}^{K}w_{k}^{2}\frac{\left(S_{k}^{2}\right)^{2}}{\nu_{k}}}.
\]

These results mean that the weights $w_{k}$ as well as the variance
component estimators $S_{k}^{2}$ can be rescaled by multiplication
with an arbitrary constant applied to all $k=1,\dots,K$ quantities.

\section{Prior Improvements }

von Davier (2025) provided an improved approximation for small component
d.f. $\nu_{k}$ and equal weights. The proposed effective d.f. is
based on the observation that for small d.f. the simple replacement
of $\sigma^{2}$ by substitution with $S^{2}$ does not hold in expectation.
That is, 
\[
E\left(\left[S_{k}^{2}\right]^{2}\right)\ne\left[\sigma_{k}^{2}\right]^{2}
\]

if $\nu_{k}<<\infty.$ For d.f. equals 1, it can be shown easily that
$E\left[\left(S^{2}\right)^{2}\right]=\sigma^{4}E\left[Z^{4}\right]=3\sigma^{4}>\sigma^{4}.$ 

This leads to a simple adjustment for the denominator
\[
B_{new}=\sum_{k=1}^{K}w_{k}^{2}\frac{\left(S_{k}^{2}\right)^{2}}{\nu_{k}+2}
\]
so that for $\nu_{k}=1$ we adjust for $E\left(Z^{4}\right)=3$ while
for $\nu_{k}\rightarrow\infty$ we have 
\[
\frac{\left(S_{k}^{2}\right)^{2}}{\nu_{k}+2}\rightarrow\frac{\left(S_{k}^{2}\right)^{2}}{\nu_{k}}.
\]
For the numerator $A,$ von Davier (2025) proposes the adjustment
\[
A_{new}=A\left(\frac{\left[K-1\right]\overline{\nu}+2}{\left[K-1\right]\overline{\nu}}\right)^{-1}=A\left(1+\frac{2}{\left[K-1\right]\overline{\nu}}\right)^{-1}=A\times f\left(\nu_{\cdot},K\right)
\]
with
\[
\overline{\nu}=\sum_{k=1}^{K}\frac{\nu_{k}}{K}
\]
being the arithmentic mean of the $\nu_{k}.$ The derivations by von
Davier (2025) provide the following expression for the unweighted
case:
\[
\nu_{*}\approx\frac{\left(\sum_{k=1}^{K}S_{k}^{2}\right)^{2}}{\left(1+\frac{K}{K-1}\frac{2}{\sum_{k}\nu_{k}}\right)\left[\sum_{k=1}^{K}\frac{\left(S_{k}^{2}\right)^{2}}{\nu_{k}+2}\right]}.
\]

\section{An Improved Effective d.f. Estimator for the Weighted Case}

Denoting the weighted average of the component d.f. $\nu_{k}$ by
\[
\overline{\nu_{w}}=\frac{\sum_{k}w_{k}\nu_{k}}{\sum w_{k}}
\]

it is proposed to utilize for cases where $K\ge2$ 
\[
A_{new}=A\times\left(1+\frac{2}{(K-1)\overline{\nu_{w}}}\right)^{-1}.
\]
The above definition yields

\begin{equation}
\nu_{*}\approx\frac{A_{new}}{B_{new}}=\frac{\left(\sum_{k=1}^{K}w_{k}S_{k}^{2}\right)^{2}}{\left(1+\frac{2}{(K-1)\overline{\nu_{w}}}\right)\left[\sum_{k=1}^{K}w_{k}^{2}\frac{\left(S_{k}^{2}\right)^{2}}{\nu_{k}+2}\right]}
\end{equation}

for a synthetic varince defined as weighted sums of variance components.
Note that 
\[
\frac{\left(\sum_{k=1}^{K}w_{k}S_{k}^{2}\right)^{2}}{\left(1+\frac{2}{(K-1)\overline{\nu_{w}}}\right)\left[\sum_{k=1}^{K}w_{k}^{2}\frac{\left(S_{k}^{2}\right)^{2}}{\nu_{k}+2}\right]}\rightarrow\frac{\left(\sum_{k=1}^{K}w_{k}S_{k}^{2}\right)^{2}}{\left[\sum_{k=1}^{K}w_{k}^{2}\frac{\left(S_{k}^{2}\right)^{2}}{\nu_{k}}\right]}
\]
as the (weighted) sum of the component degrees of freedom $\sum_{k}w_{k}\nu_{k}\rightarrow\infty$
grows. Similarly, with $K\rightarrow\infty$, we have that 
\[
\left(1+\frac{2}{(K-1)\overline{\nu_{w}}}\right)\rightarrow1
\]
 even if $\nu_{k}=1$ remains for all $k.$ 

\section{Further Improvement of the Approximation}

The adjustment factor 
\[
\frac{3}{\left(1+\frac{C}{\left[K-p\right]\overline{\nu}}\right)}
\]
is a simple multiplicative term that is obtained if all component
degrees of freedom are equal to one. Then we can write
\[
\nu_{*,2025}=\frac{3}{\left(1+\frac{C}{\left[K-p\right]\overline{\nu}}\right)}\nu_{*,Satt}
\]
with $p=1$ and assuming $\forall k:\nu_{k}=1=\nu$. This adjustment
was derived based on the observation that $E\left(Z^{4}\right)=3$
and on the assumption that that for $K=2$ and $\nu_{1}=\nu_{2}=1$
the maximum allowable approximate d.f. is $\nu_{*}=2.$ This maximum
is obtained only if $S_{1}^{2}=S_{2}^{2}$, i.e., if the two variance
components are identical (i.e., if they are perfectly correlated,
not independent). However, the assumption under which the Satterthwaite
equation was derived is that the $S_{1}^{2},S_{2}^{2}$ are independent. 

Also note that for the $K=2$ variance components case, we find 
\[
\frac{C_{1}}{\left[K-1\right]\overline{\nu}}=\frac{C_{2}}{K\overline{\nu}}
\]
by means of setting $C_{2}=\frac{C_{1}K}{K-1}$. That is why we can
focus on $p=0$ going forward. 

Therefore, we will only consider adjustments of the type
\[
\hat{\nu}_{*,C}=\left(1+\frac{C}{K\overline{\nu}}\right)^{-1}\frac{\left[\sum_{k=1}^{K}w_{k}S_{k}^{2}\right]^{2}}{\sum_{k=1}^{K}\frac{S_{k}^{4}}{\nu_{k}+2}}
\]
in the following. 

\subsection{Matching the Expected Value for the Minimal Case}

The case of $K=2$ and $\nu_{1}=\nu_{2}=1$ remains our minimal case
of reference for the adjustment, as it marks the smallest possible
non-trivial combination of variance components. It can be shown that
\[
1\le\frac{\left(S_{1}^{2}+S_{2}^{2}\right)^{2}}{S_{2}^{4}+S_{2}^{4}}\le2
\]
for any combination of $S_{1}^{2},S_{2}^{2}$. The true mean of the
distribution must therefor be between $1$ and $2,$ so that we know
the absolute difference is 
\[
\left|E\left[\frac{\left(S_{1}^{2}+S_{2}^{2}\right)^{2}}{S_{2}^{4}+S_{2}^{4}}\right]-1.5\right|\le0.5.
\]
Picking the midpoint, $M=1.5$ for determining the correction constant
$C,$we find
\[
2=\frac{3}{\left(1+\frac{C}{2}\right)}1.5\leftrightarrow2\left(1+\frac{C}{2}\right)=4.5\leftrightarrow C=2.25
\]
as a value that should provide an average of the approximate d.f.
of $M\left(\hat{\nu}_{*}\right)\approx2.00$ for $K=2,\nu=1.$ 

That means, using these considerations, one could propose the following
adjustment to the Satterthwaite (1941, 1946) equation:
\[
\hat{\nu}_{*,2.25}=\frac{\left(\sum_{k}w_{k}S_{k}^{2}\right)^{2}}{\left(1+\frac{2.25}{K\overline{\nu}}\right)\left(\sum_{k}w_{k}^{2}\frac{\left[S_{k}^{2}\right]^{2}}{\nu_{k}+2}\right)}
\]
which retains the same limiting behavior for $K\rightarrow\infty$
and $\nu_{k}\rightarrow\infty$. For these limits, we have 
\[
\hat{\nu}_{*,2.25}\rightarrow\frac{\left(\sum_{k}w_{k}S_{k}^{2}\right)^{2}}{\left(\sum_{k}w_{k}^{2}\frac{\left[S_{k}^{2}\right]^{2}}{\nu_{k}}\right)}=\hat{\nu}_{*,Satt}.
\]

It could be argued that a rational strategy is to adjust the equation
focusing on the case with $\nu=1$ and $K=2$ in an effort to match
the observed average. When looking at the unadjusted estimator, a
simulation of 4 million replications gives an observed average value
of 
\[
E\left(\frac{\left[Z_{1}^{2}+Z_{2}^{2}\right]^{2}}{Z_{1}^{4}+Z_{2}^{4}}\right)\approx\frac{1}{4,000,000}\sum_{i=1}^{4,000,000}\frac{\left[Z_{1}^{2}+Z_{2}^{2}\right]^{2}}{Z_{1}^{4}+Z_{2}^{4}}=1.41425.
\]
This could be used to find a correction term $C$ for which 
\[
E\left(\nu_{*}\right)=2=\left(\frac{3}{1+\frac{C}{2}}\right)1.41425
\]

this leads to 
\[
C=\frac{6-2\times1.41425}{1.41425}\approx2.24,
\]
a number that is (surprisingly???) close to the value $2.25$ obtained
using a simple approximation. When using this adjustment $\left(1+\frac{2.2425}{K\overline{\nu}}\right)^{-1}$,
we obtain the following table \ref{tab: exact 2.0} that indeed shows
the average estimated d.f. matches the expected value for the minimal
case $K=2,\nu=1$ closely, we have $E\left(\nu_{*}\right)=M\left(\hat{\nu}_{K=2,\nu=1}\right)=2.00$. 

\begin{figure}
\includegraphics[scale=0.4]{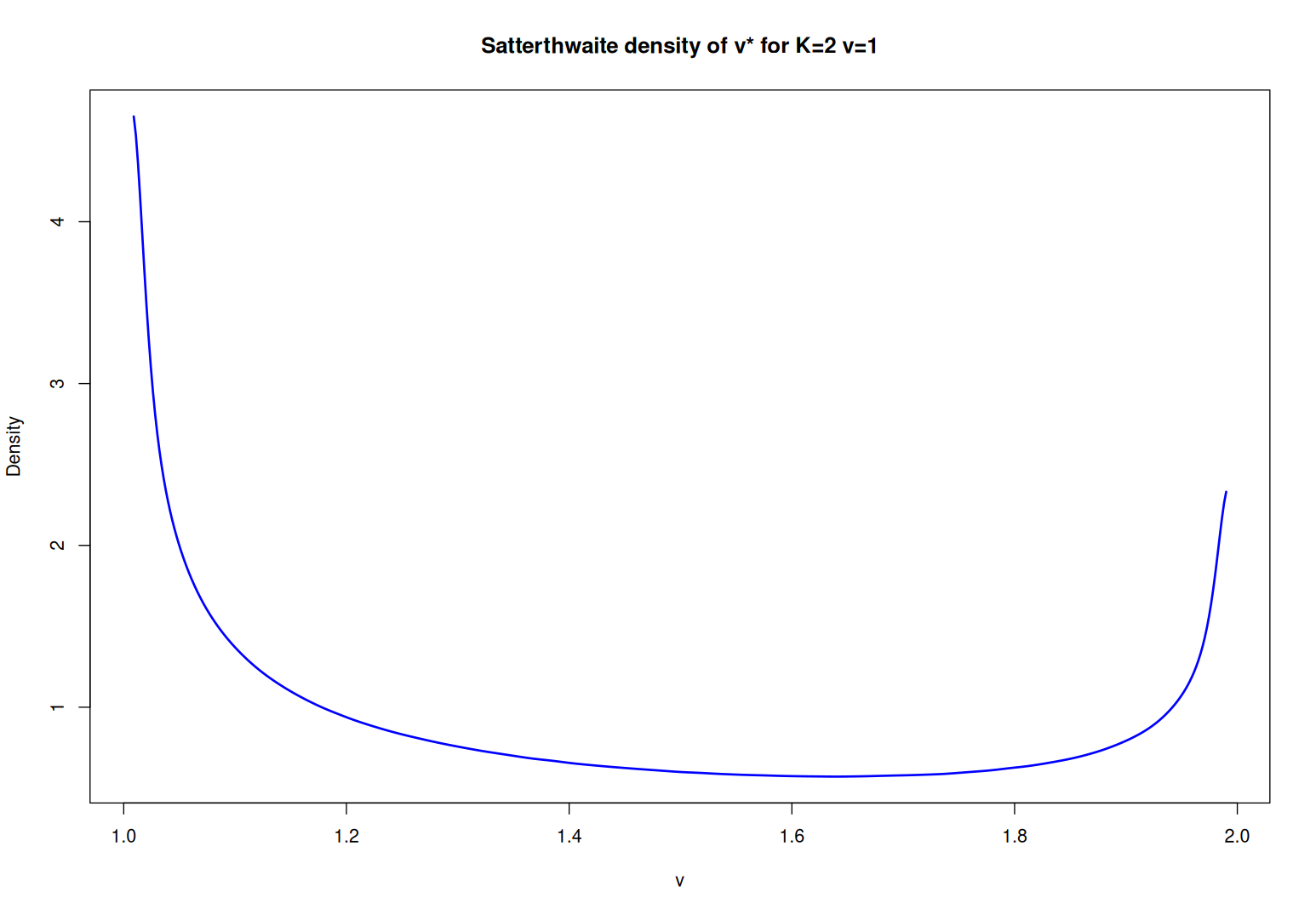}

\caption{Empirical density of $R$ based on 4,000,000 replicates of two independent
normal deviates.}\label{fig:Empirical-density}
\end{figure}

However, two things to note here. First, this adjustment focuses on
one case that does not seem to be very common, a case where there
are only two components is not rare, of course, but cases where only
two variance components are added together, and each component is
essentially a squared normallly distributed variable, with one degree
of freedom $\nu=1,$seem to be not those seen out in the wild often.
Second, the behavior of the expression 
\[
\frac{\left[Z_{1}^{2}+Z_{2}^{2}\right]^{2}}{Z_{1}^{4}+Z_{2}^{4}}=\hat{\nu}_{satt}
\]
for two independent normal deviates is shown in figure \ref{fig:Empirical-density}.
It is obvious that the most frequently observed values are close to
1 and 2. This is the case because $Z_{1},Z_{2}$ are independent and
in the more likely case that $Z_{i}<<Z_{j}$ then $\hat{\nu}_{satt}\approx1$
while it is less likely that $Z_{1}\approx Z_{2}$ which leads to
$\hat{\nu}_{satt}\approx2$. So it would seem that focusing on a value
that matches the empirical average of $1.41425$ for a single $K,\nu$
combinations is not the most adviseable strategy.

This lead to a consideration that goes beyond finding an adjustment
that exactly matches the theoretical value for one of the potential
combinations of $K,\nu$. Instead, the next section explores how a
constant can be found that minimizes the difference from the expected
result $K\nu$ for a range of components and component degrees of
freedom.

\subsection{Optimal Adjustment for a Range of Cases}

In this section, we will step away from the somewhat reductionist
view that limits the problem to only matching the expected value $E\left(\nu_{*}\right)=K\nu$
for $K=2$ and equal component degrees of freedom. Instead, this section
looks at the small d.f. behavior of a range of adjustment factors
$C\in\Omega_{C}=\left]2.0,3.2\right[$(why we do not look outside
that range will be clear when examining figure ). We aim to cover
the small d.f. and low component count cases where the adjustment
to the effective d.f. are expected to have the largest effects. The
simulations use the best case scenario where we know each component
is $\chi^{2}$ distributed with d.f. $\nu_{k}$, that is, each variance
component can be written as
\[
S_{k}^{2}=\sum_{j=1}^{\nu_{k}}Z_{j}^{2}
\]
with $Z_{j}\sim N\left(0,1\right)$, so that $E\left(S_{k}^{2}\right)=\nu_{k}$
and $Var\left(S_{k}^{2}\right)=2\nu_{k}$. In this case, for $S_{*}^{2}=\sum_{k=1}^{K}S_{k}^{2}$
we can determine the expected value is $E\left(\nu_{*}\right)=K\times\nu$.
Within the range of potential constants $C\in\Omega_{C}$, the choice
will be made based on an optimality criterion
\[
X_{C}^{2}=\sum_{K\in\Omega_{K}}\sum_{\nu\in\Omega_{\nu}}\frac{\left[M\left(\hat{\nu}_{*,C}\right)-K\nu\right]^{2}}{K\nu}
\]
where the range of components $K$ and component degrees of freedom
$\nu$ is defined as 

\[
K\in\Omega_{K}=\left\{ 2,3,\dots,K_{max}\right\} \,\,\mathrm{and}\,\,\nu\in\Omega_{\nu}=\left\{ 1,2,\dots,\nu_{max}\right\} .
\]

The quantity $X_{C}^{2}$ can be understood as a pseudo $\chi^{2}$
measure of deviation, which is minimized by finding 
\[
C_{opt}=\arg\min_{C\in\Omega_{C}}\left\{ X_{C}^{2}\right\} 
\]
to determine an optimal choice for this range of components $K$ and
compoment d.f. $\nu$. In order to obtain estimates of the expected
effective d.f. $E\left(\nu_{*,C}\right)$ a number of $r=10000$ replicates
was used to calculate 
\[
\frac{1}{10000}\sum_{i=1}^{10000}\hat{\nu}_{*,C}=M\left(\hat{\nu}_{*,C}\right)\approx E\left(\nu_{*,C}\right).
\]

In terms of the practical implementation of the search, the optimal
constant $C_{opt}$ was determined using a line search for different
values of $\left(K_{max},\nu_{max}\right)$ and it was monitored if
the choice converges for these different upper limits. More specifically,
the following combinations were checked:

\[
\left(K_{max},\nu_{max}\right)\in\left\{ \left(5,5\right),\left(10,10\right),\left(20,20\right),\dots,\left(80,80\right),\left(100,100\right)\right\} 
\]

Then, $C_{opt}$ is used to define the improved effective degree of
freedom as 
\begin{equation}
\nu_{*}\approx\left(1+\frac{C_{opt}}{K\overline{\nu_{w}}}\right)^{-1}\frac{\left(\sum_{k=1}^{K}w_{k}S_{k}^{2}\right)^{2}}{\sum_{k=1}^{K}\frac{w_{k}^{2}S_{k}^{4}}{\nu_{k}+2}}.\label{eq:new_adjust}
\end{equation}

Note that, as before, the correction factor $\left(1+\frac{C_{opt}}{K\overline{\nu}}\right)^{-1}$is
only important for small $K,\overline{\nu}$ as the limiting case
is obtained when assuming 
\[
K,\overline{\nu}\rightarrow\infty
\]
with $\nu_{k}=\nu$ for all $k=1,\dots,K$. In that case we find the
well known result
\[
\lim_{K,\nu\rightarrow\infty}\left(1+\frac{C}{K\nu}\right)^{-1}\left[\nu+2\right]\frac{\left(\sum_{k}S_{k}^{2}\right)^{2}}{\sum S_{k}^{4}}=\nu\frac{K^{2}\sigma^{4}}{K\sigma^{4}}=K\nu.
\]
That is why it is sufficient to examine a set of limited ranges $\Omega_{K}\times\Omega_{\nu}=\left\{ 2,3,\dots,K_{max}\right\} \times\left\{ 1,2,\dots,\nu_{max}\right\} $
as with growing $K,\nu$ the adjustment will converge to the original
Satterthwaite (1946) equation. 

\section{Empirical Evidence for the New Adjustment }

This section explains how the value $C_{opt}$ was chosen to provide
an optimal adjustment for a range of small to medium values for $K$
and $\nu_{k}$ and how the improvement of the Satterthwaite (1941,
1946) effective degrees of freedom fares compared to the original,
and in comparison to the improvement proposed by von Davier (2025)
as well as the contender that aims to match the expected degrees of
freedom using $C=2.25,$to the theoretical value for $K=2$ and $\nu_{1}=\nu_{2}$=1. 

\subsection{Determining $C_{opt}$}

\begin{figure}
\includegraphics[scale=0.7]{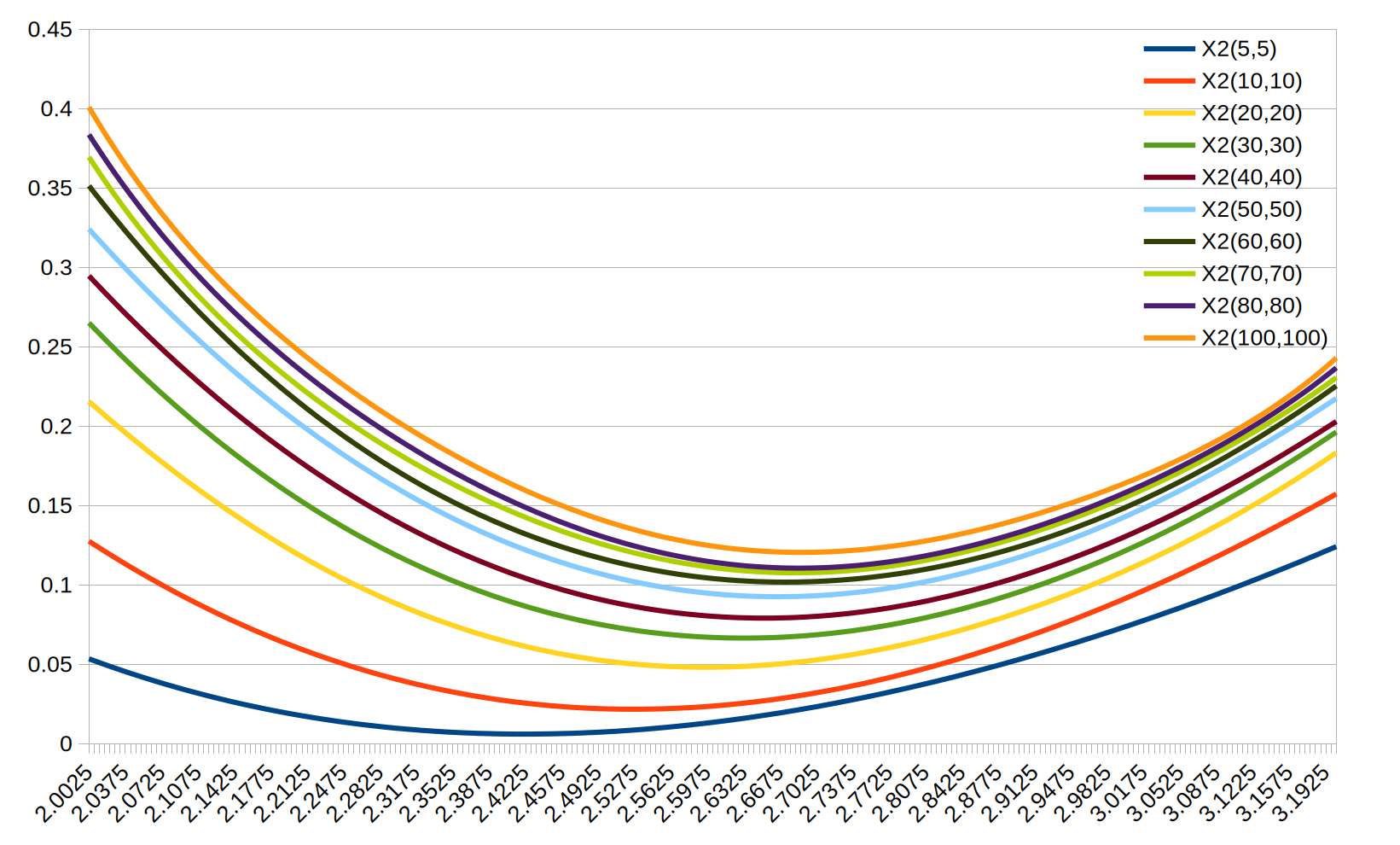}

\caption{ Observed $X^{2}$ discrepancies for different values of $K_{max},\nu_{max}$}\label{fig: Opt_C_fig}
\end{figure}

Figure \ref{fig: Opt_C_fig} shows the graph for $X^{2}$ as a function
of $C$ for the different selections of $K_{max},\nu_{max}.$ This
graph was produced by smoothing the data using a polynomous regression
with an order of up to 6. The order of the polynome was determined
based on minimising the RMSE using a 10-fold cross validation.

It appears that the minimum value starts out at around $C_{opt}\left(5,5\right)\approx2.42$
and converges as $K_{max}$ and $\nu_{max}$ increase to about $2.66\le C_{opt}\le2.69$.

The way the optimal value was determined as $C_{opt}\approx2.69$
can be explained based on table \ref{tab:growth_precise}. The growth
of the values of $C_{opt}$ as $K_{max},\nu_{max}$ grow is negatively
accelerated and for the difference between $\left(40,40\right)$ and
$\left(50,50\right)$ we see that $C_{opt}$only increases by $0.49%\%
$ from $2.6524$ to $2.6655.$The percentage increase seems to be cut
in half for each increase of $K_{max},\nu_{max}$ by 10. Extrapolating
this pattern to the increas from $\left(80,80\right)$ to $\left(100,100\right)$
{[}which is a table of $10000$ cases{]} shows only an increase by
$0.0078\%$ that implies a limited growth of $C_{opt}$ as $K_{max},\nu_{max}$
grows. We therefore assert that $C_{opt}\approx2.69$ can be chosen
as an optimal value.

\begin{table}
\selectlanguage{english}%
\begin{tabular}{|c|c|c|c|c|c|}
\hline 
$K_{max},\nu_{max}$ & Cells in $X^{2}$ & \textbf{Degree} & \textbf{$R^{2}$} & \textbf{Opt $C$} & \textbf{Min $X^{2}$}\tabularnewline
\hline 
\hline 
$\left(5,5\right)$ & 25 & 4 & 0.9989 & 2.4211 & 0.0062\tabularnewline
\hline 
\foreignlanguage{american}{$\left(10,10\right)$} & 100 & 3 & 0.9979 & 2.5263 & 0.0218\tabularnewline
\hline 
\foreignlanguage{american}{$\left(20,20\right)$} & 400 & 5 & 0.9978 & 2.5954 & 0.0483\tabularnewline
\hline 
\foreignlanguage{american}{$\left(30,30\right)$} & 900 & 4 & 0.9983 & 2.6329 & 0.0666\tabularnewline
\hline 
\foreignlanguage{american}{$\left(40,40\right)$} & 1600 & 3 & 0.9978 & 2.6524 & 0.0792\tabularnewline
\hline 
\foreignlanguage{american}{$\left(50,50\right)$} & 2500 & 4 & 0.9911 & 2.6655 & 0.0927\tabularnewline
\hline 
\foreignlanguage{american}{$\left(60,60\right)$} & 3600 & 4 & 0.9917 & 2.6727 & 0.1019\tabularnewline
\hline 
\foreignlanguage{american}{$\left(70,70\right)$} & 4900 & 6 & 0.9918 & 2.6824 & 0.1078\tabularnewline
\hline 
\foreignlanguage{american}{$\left(80,80\right)$} & 6400 & 6 & 0.9925 & 2.6858 & 0.1107\tabularnewline
\hline 
\foreignlanguage{american}{$\left(100,100\right)$} & 10000 & 6 & 0.9934 & 2.6879 & 0.1206\tabularnewline
\hline 
\end{tabular}
\centering{}\caption{Optimal values and growth rates for $C_{opt}$ by different combinations
of $K_{max},\nu_{max}$ranging from $(5,5)$ to $(100,100)$. }\label{tab:growth_precise}
\selectlanguage{american}%
\end{table}

\subsection{Comparing the Adjustments}

The following tables show the averages for the unadjusted $\hat{\nu}_{*,Satter}$,
for the von Davier (2025) adjustment $\hat{\nu}_{*,vD}$, and for
the proposed improved adjustment using $C_{opt}=2.69,$ as well as
for $\hat{\nu}_{*,2.25}$ to match the $K=2,\nu=1$ case exactly.
The following tables show results for the average estimates of $\hat{\nu}_{*}$over
10000 replications and for the following subset of component counts
and component degrees of freedom

\[
K\in\Omega_{K/2}=\left\{ 2,4,6,8,10,20,40,160\right\} 
\]

and 
\[
\nu\in\Omega_{\nu/2}=\left\{ 1,3,5,7,9,15,30,80\right\} 
\]

This selection was chosen make it possible to show tables on a single
page. Their associated $X^{2}$ sums of standardized squared deviations
for each case are given in an additional table based on the full set
of $K,\nu$ combinations. The results in tables\ref{tab:Averages original Satter},\ref{tab: old improve 2025},\ref{tab: exact 2.0}
and \ref{tab:Averages C 2.8} are based on 10000 replications each
and the average in the table can be compared to the product of the
row and column header value.

In the tables, the average of the estimated d.f. are shown. The columns
specify the degrees of freedom $\nu$ per component, the rows specify
the number of components $K$. First, the original Satterthwaite (1941,
1946) formula is shown. It is known that this estimator of the effective
degrees of freedom is biased when the component degrees of freedom
are small. 

\begin{table}
{\scriptsize{}%
\begin{tabular}{|c|r|r|r|r|r|r|r|r|}
\hline 
$K\backslash\nu$ & 1 & 3 & 5 & 7 & 9 & 15 & 30 & 80\tabularnewline
\hline 
\hline 
\foreignlanguage{english}{2} & \foreignlanguage{english}{1.42} & \foreignlanguage{english}{4.98} & \foreignlanguage{english}{8.77} & \foreignlanguage{english}{12.63} & \foreignlanguage{english}{16.53} & \foreignlanguage{english}{28.40} & \foreignlanguage{english}{58.22} & \foreignlanguage{english}{158.10}\tabularnewline
\hline 
\foreignlanguage{english}{4} & \foreignlanguage{english}{2.19} & \foreignlanguage{english}{8.79} & \foreignlanguage{english}{16.06} & \foreignlanguage{english}{23.67} & \foreignlanguage{english}{31.37} & \foreignlanguage{english}{54.98} & \foreignlanguage{english}{114.50} & \foreignlanguage{english}{314.24}\tabularnewline
\hline 
\foreignlanguage{english}{6} & \foreignlanguage{english}{2.93} & \foreignlanguage{english}{12.45} & \foreignlanguage{english}{23.36} & \foreignlanguage{english}{34.67} & \foreignlanguage{english}{46.20} & \foreignlanguage{english}{81.49} & \foreignlanguage{english}{170.80} & \foreignlanguage{english}{470.27}\tabularnewline
\hline 
\foreignlanguage{english}{8} & \foreignlanguage{english}{3.66} & \foreignlanguage{english}{16.14} & \foreignlanguage{english}{30.54} & \foreignlanguage{english}{45.51} & \foreignlanguage{english}{60.94} & \foreignlanguage{english}{107.98} & \foreignlanguage{english}{227.04} & \foreignlanguage{english}{626.48}\tabularnewline
\hline 
\foreignlanguage{english}{10} & \foreignlanguage{english}{4.36} & \foreignlanguage{english}{19.86} & \foreignlanguage{english}{37.77} & \foreignlanguage{english}{56.43} & \foreignlanguage{english}{75.70} & \foreignlanguage{english}{134.34} & \foreignlanguage{english}{283.21} & \foreignlanguage{english}{782.49}\tabularnewline
\hline 
\foreignlanguage{english}{20} & \foreignlanguage{english}{7.81} & \foreignlanguage{english}{37.96} & \foreignlanguage{english}{73.39} & \foreignlanguage{english}{110.88} & \foreignlanguage{english}{149.31} & \foreignlanguage{english}{266.92} & \foreignlanguage{english}{564.55} & \foreignlanguage{english}{1563.00}\tabularnewline
\hline 
\foreignlanguage{english}{40} & \foreignlanguage{english}{14.63} & \foreignlanguage{english}{74.08} & \foreignlanguage{english}{145.17} & \foreignlanguage{english}{219.90} & \foreignlanguage{english}{296.66} & \foreignlanguage{english}{531.68} & \foreignlanguage{english}{1127.06} & \foreignlanguage{english}{3124.05}\tabularnewline
\hline 
\foreignlanguage{english}{160} & \foreignlanguage{english}{54.79} & \foreignlanguage{english}{290.21} & \foreignlanguage{english}{573.48} & \foreignlanguage{english}{873.04} & \foreignlanguage{english}{1180.50} & \foreignlanguage{english}{2119.89} & \foreignlanguage{english}{4502.64} & \foreignlanguage{english}{12489.85}\tabularnewline
\hline 
\end{tabular}}{\scriptsize\par}

\caption{Averages of the estimated effective d.f. using the original Satterthwaite
(1941, 1946) approach.}\label{tab:Averages original Satter}
\end{table}

For the original Satterthwaite (1941, 1946) approach, the empirical
averages are given in table \ref{tab:Averages original Satter}. It
can be seen that for component d.f. $\nu\in\left\{ 1,3,5,7,9\right\} $
the estimated effective d.f. $\hat{\nu}_{*}$are much smaller than
the true value, that is, $\hat{\nu}_{*,Satter}<<K\times\nu$. For
example, the true value for $K=6$ and $\nu=3$ is $E\left(\nu_{*}\right)=18$
while the observed average is $M\left(\hat{\nu}_{*}\right)=12.45.$
Even for $\nu=9$ the estimator does not closely track the true value,
for $K=4$ we find $M\left(\hat{\nu}_{*}\right)=31.37<36.$ Maybe
surprisingly, even for $\nu=30,K=160$ we obtain $M\left(\hat{\nu}_{*}\right)=4502.64<4800=30\times160.$ 

The following table \ref{tab: old improve 2025} contains the results
for the adjustment suggested by von Davier (2025), where $\left(1+\frac{2}{\left(K-1\right)\overline{\nu}}\right)^{-1}$
was proposed. The differences are small, but it appears that the new
adjustment using $C_{opt}$ is closer to the expected values by some
amount. 

\begin{table}
{\scriptsize{}%
\begin{tabular}{|c|r|r|r|r|r|r|r|r|}
\hline 
$K\backslash\nu$ & 1 & 3 & 5 & 7 & 9 & 15 & 30 & 80\tabularnewline
\hline 
\hline 
\foreignlanguage{english}{2} & \foreignlanguage{english}{1.42} & \foreignlanguage{english}{4.97} & \foreignlanguage{english}{8.77} & \foreignlanguage{english}{12.61} & \foreignlanguage{english}{16.55} & \foreignlanguage{english}{28.36} & \foreignlanguage{english}{58.20} & \foreignlanguage{english}{158.04}\tabularnewline
\hline 
\foreignlanguage{english}{4} & \foreignlanguage{english}{3.95} & \foreignlanguage{english}{11.95} & \foreignlanguage{english}{19.88} & \foreignlanguage{english}{27.76} & \foreignlanguage{english}{35.71} & \foreignlanguage{english}{59.68} & \foreignlanguage{english}{119.49} & \foreignlanguage{english}{319.42}\tabularnewline
\hline 
\foreignlanguage{english}{6} & \foreignlanguage{english}{6.28} & \foreignlanguage{english}{18.35} & \foreignlanguage{english}{30.15} & \foreignlanguage{english}{42.11} & \foreignlanguage{english}{54.06} & \foreignlanguage{english}{89.99} & \foreignlanguage{english}{179.74} & \foreignlanguage{english}{479.62}\tabularnewline
\hline 
\foreignlanguage{english}{8} & \foreignlanguage{english}{8.48} & \foreignlanguage{english}{24.58} & \foreignlanguage{english}{40.46} & \foreignlanguage{english}{56.34} & \foreignlanguage{english}{72.19} & \foreignlanguage{english}{120.17} & \foreignlanguage{english}{239.90} & \foreignlanguage{english}{639.62}\tabularnewline
\hline 
\foreignlanguage{english}{10} & \foreignlanguage{english}{10.71} & \foreignlanguage{english}{30.75} & \foreignlanguage{english}{50.51} & \foreignlanguage{english}{70.34} & \foreignlanguage{english}{90.30} & \foreignlanguage{english}{150.06} & \foreignlanguage{english}{300.02} & \foreignlanguage{english}{799.78}\tabularnewline
\hline 
\foreignlanguage{english}{20} & \foreignlanguage{english}{21.23} & \foreignlanguage{english}{61.00} & \foreignlanguage{english}{100.87} & \foreignlanguage{english}{140.91} & \foreignlanguage{english}{180.51} & \foreignlanguage{english}{300.37} & \foreignlanguage{english}{600.03} & \foreignlanguage{english}{1599.88}\tabularnewline
\hline 
\foreignlanguage{english}{40} & \foreignlanguage{english}{41.71} & \foreignlanguage{english}{121.32} & \foreignlanguage{english}{200.89} & \foreignlanguage{english}{280.83} & \foreignlanguage{english}{360.65} & \foreignlanguage{english}{600.35} & \foreignlanguage{english}{1200.25} & \foreignlanguage{english}{3199.69}\tabularnewline
\hline 
\foreignlanguage{english}{160} & \foreignlanguage{english}{162.21} & \foreignlanguage{english}{481.46} & \foreignlanguage{english}{801.11} & \foreignlanguage{english}{1120.34} & \foreignlanguage{english}{1440.82} & \foreignlanguage{english}{2400.28} & \foreignlanguage{english}{4800.55} & \foreignlanguage{english}{12800.27}\tabularnewline
\hline 
\end{tabular}}{\scriptsize\par}

\caption{Averages of the estimated effective d.f. for the von Davier (2025)
adjustment using $C=2$ and $K-1$}\label{tab: old improve 2025}
\end{table}

It can be seen that the expected value for $K=2$ and $\nu=1$ is
the same as for the original Satterthwaite equation, but otherwise
the adjustment proposed by von Davier (2025) seems to track the theoretical
values of $K\times\nu$ closely. However, the next two tables will
show that closer approximations are possible. 

The next table shows the adjustment for $C=2.24$ which attempts to
match $K\nu=2\times1$ closely for the limiting case with only two
components and single d.f. It can be seen that the expected value
for this adjusted Satterthwaite formula is indeed $2.00$ as given
in table \ref{tab: exact 2.0}.

\begin{table}
{\scriptsize{}%
\begin{tabular}{|c|r|r|r|r|r|r|r|r|}
\hline 
$K\backslash\nu$ & 1 & 3 & 5 & 7 & 9 & 15 & 30 & 80\tabularnewline
\hline 
\hline 
\foreignlanguage{english}{2} & \foreignlanguage{english}{2.00} & \foreignlanguage{english}{6.05} & \foreignlanguage{english}{10.01} & \foreignlanguage{english}{13.99} & \foreignlanguage{english}{18.00} & \foreignlanguage{english}{29.93} & \foreignlanguage{english}{59.83} & \foreignlanguage{english}{159.83}\tabularnewline
\hline 
\foreignlanguage{english}{4} & \foreignlanguage{english}{4.22} & \foreignlanguage{english}{12.28} & \foreignlanguage{english}{20.23} & \foreignlanguage{english}{28.21} & \foreignlanguage{english}{36.09} & \foreignlanguage{english}{60.06} & \foreignlanguage{english}{119.97} & \foreignlanguage{english}{319.78}\tabularnewline
\hline 
\foreignlanguage{english}{6} & \foreignlanguage{english}{6.41} & \foreignlanguage{english}{18.53} & \foreignlanguage{english}{30.32} & \foreignlanguage{english}{42.21} & \foreignlanguage{english}{54.20} & \foreignlanguage{english}{90.03} & \foreignlanguage{english}{179.89} & \foreignlanguage{english}{479.87}\tabularnewline
\hline 
\foreignlanguage{english}{8} & \foreignlanguage{english}{8.51} & \foreignlanguage{english}{24.59} & \foreignlanguage{english}{40.46} & \foreignlanguage{english}{56.36} & \foreignlanguage{english}{72.22} & \foreignlanguage{english}{120.09} & \foreignlanguage{english}{239.94} & \foreignlanguage{english}{639.94}\tabularnewline
\hline 
\foreignlanguage{english}{10} & \foreignlanguage{english}{10.72} & \foreignlanguage{english}{30.62} & \foreignlanguage{english}{50.56} & \foreignlanguage{english}{70.44} & \foreignlanguage{english}{90.31} & \foreignlanguage{english}{150.17} & \foreignlanguage{english}{299.88} & \foreignlanguage{english}{799.80}\tabularnewline
\hline 
\foreignlanguage{english}{20} & \foreignlanguage{english}{21.11} & \foreignlanguage{english}{60.89} & \foreignlanguage{english}{100.71} & \foreignlanguage{english}{140.50} & \foreignlanguage{english}{180.28} & \foreignlanguage{english}{300.02} & \foreignlanguage{english}{600.16} & \foreignlanguage{english}{1600.04}\tabularnewline
\hline 
\foreignlanguage{english}{40} & \foreignlanguage{english}{41.63} & \foreignlanguage{english}{121.18} & \foreignlanguage{english}{200.69} & \foreignlanguage{english}{280.73} & \foreignlanguage{english}{360.63} & \foreignlanguage{english}{600.30} & \foreignlanguage{english}{1199.88} & \foreignlanguage{english}{3199.90}\tabularnewline
\hline 
\foreignlanguage{english}{160} & \foreignlanguage{english}{161.96} & \foreignlanguage{english}{481.45} & \foreignlanguage{english}{801.08} & \foreignlanguage{english}{1120.87} & \foreignlanguage{english}{1440.76} & \foreignlanguage{english}{2400.07} & \foreignlanguage{english}{4800.32} & \foreignlanguage{english}{12799.96}\tabularnewline
\hline 
\end{tabular}}{\scriptsize\par}

\caption{Averages of the estimated effective d.f. for the adjustment using
$C=2.24$ and $K$}\label{tab: exact 2.0}
\end{table}

Table \ref{tab:Averages C 2.8}shows results for the proposed improved
adjustment using the factor $\left(1+\frac{2.69}{K\overline{\nu}}\right)^{-1}$
according to equation \ref{eq:new_adjust}. As an example, for $\nu=9$
and $K=6$ we expect to see $E\left(\nu_{*}\right)=9\times6=54$,
and indeed we observe $M\left(\hat{\nu}_{*}\right)=53.76\approx54$,
and for $\nu=3,K=8$ we obtain $M\left(\hat{\nu}_{*}\right)=24.11\approx24$.

\begin{table}
{\scriptsize{}%
\begin{tabular}{|c|r|r|r|r|r|r|r|r|}
\hline 
$K\backslash\nu$ & 1 & 3 & 5 & 7 & 9 & 15 & 30 & 80\tabularnewline
\hline 
\hline 
\foreignlanguage{english}{2} & \foreignlanguage{english}{1.80 } & \foreignlanguage{english}{5.71 } & \foreignlanguage{english}{9.67 } & \foreignlanguage{english}{13.61 } & \foreignlanguage{english}{17.56 } & \foreignlanguage{english}{29.53 } & \foreignlanguage{english}{59.44 } & \foreignlanguage{english}{159.38 }\tabularnewline
\hline 
\foreignlanguage{english}{4} & \foreignlanguage{english}{3.93 } & \foreignlanguage{english}{11.92 } & \foreignlanguage{english}{19.88 } & \foreignlanguage{english}{27.76 } & \foreignlanguage{english}{35.71 } & \foreignlanguage{english}{59.66 } & \foreignlanguage{english}{119.46 } & \foreignlanguage{english}{319.33 }\tabularnewline
\hline 
\foreignlanguage{english}{6} & \foreignlanguage{english}{6.05 } & \foreignlanguage{english}{18.09 } & \foreignlanguage{english}{29.94 } & \foreignlanguage{english}{41.93 } & \foreignlanguage{english}{53.74 } & \foreignlanguage{english}{89.57 } & \foreignlanguage{english}{179.53 } & \foreignlanguage{english}{479.26 }\tabularnewline
\hline 
\foreignlanguage{english}{8} & \foreignlanguage{english}{8.23 } & \foreignlanguage{english}{24.09 } & \foreignlanguage{english}{40.09 } & \foreignlanguage{english}{55.99 } & \foreignlanguage{english}{71.85 } & \foreignlanguage{english}{119.68 } & \foreignlanguage{english}{239.55 } & \foreignlanguage{english}{639.31 }\tabularnewline
\hline 
\foreignlanguage{english}{10} & \foreignlanguage{english}{10.36 } & \foreignlanguage{english}{30.37 } & \foreignlanguage{english}{50.10 } & \foreignlanguage{english}{69.99 } & \foreignlanguage{english}{89.82 } & \foreignlanguage{english}{149.69 } & \foreignlanguage{english}{299.41 } & \foreignlanguage{english}{799.46 }\tabularnewline
\hline 
\foreignlanguage{english}{20} & \foreignlanguage{english}{20.76 } & \foreignlanguage{english}{60.47 } & \foreignlanguage{english}{100.04 } & \foreignlanguage{english}{139.89 } & \foreignlanguage{english}{180.16 } & \foreignlanguage{english}{299.44 } & \foreignlanguage{english}{599.75 } & \foreignlanguage{english}{1599.48 }\tabularnewline
\hline 
\foreignlanguage{english}{40} & \foreignlanguage{english}{41.12 } & \foreignlanguage{english}{120.55 } & \foreignlanguage{english}{200.39 } & \foreignlanguage{english}{280.16 } & \foreignlanguage{english}{359.85 } & \foreignlanguage{english}{599.66 } & \foreignlanguage{english}{1199.46 } & \foreignlanguage{english}{3199.37 }\tabularnewline
\hline 
\foreignlanguage{english}{160} & \foreignlanguage{english}{161.96 } & \foreignlanguage{english}{481.15 } & \foreignlanguage{english}{800.70 } & \foreignlanguage{english}{1119.76 } & \foreignlanguage{english}{1440.00 } & \foreignlanguage{english}{2399.63 } & \foreignlanguage{english}{4799.90 } & \foreignlanguage{english}{12798.99 }\tabularnewline
\hline 
\end{tabular}}{\scriptsize\par}

\caption{Averages of the estimated effective d.f. for the adjustment using
$C=2.69$ and $K$}\label{tab:Averages C 2.8}
\end{table}

It can be seen that the observed average of the estimated degrees
of freedom $M\left(\hat{\nu}_{*}\right)$ closely track the theoretical
value $K\times\nu$ for most entries in this table. Especially for
$\nu=1$, the cases shown in the first column, the new adjustment
provides a closer tracking of the expected $E\left(\nu_{*,new}\right)=K$,
but also for varying $\nu$ and $K=2$ the new adjustment provides
a slightly better approximation than the one using $C=2.25$.

The overall quality of the approximations using different adjustments
can be evaluated using the pseudo $\chi^{2}$ statistics $X^{2}$
calculated for each of the adjustment constants $C\in\left\{ \left(2.69,K\right),\left(2,K-1\right),\left(2.25,K\right)\right\} $
as well as the original Satterthwaite (1941, 1946) formula. Note that
larger deviations can typically be expected for small $K$ and small
component $\nu$, and for large values of these the approximaytion
is quite close, so that extending the table further beyond $K>160$
and $\nu>80$ would likely not add much to the pseudo $X^{2}$ measure.
As expected, the table \ref{tab: Pseudo X2 values}shows that proposed
adjustment is indeed the one that shows the smallest discrepancy among
the four.

\begin{table}
\begin{centering}
\begin{tabular}{|r|r|l|r|}
\hline 
 & $C$ & $K_{adj}$ & $X^{2}$\tabularnewline
\hline 
\hline 
Satterthwaite (1941,1946) & N/A & N/A & 13.27251\tabularnewline
\hline 
von Davier (2025) & 2 & K-1 & 0.31631\tabularnewline
\hline 
Assume $E\left(\hat{\nu}_{*}\right)\approx1.5$ for $(K=2,\nu=1)$ & 2.25 & K & 0.06412\tabularnewline
\hline 
\textbf{\textit{Optimal $\left(K_{max},\nu_{max}\right)$ Adjustment
$C_{opt}$}} & \textbf{\textit{2.69}} & \textbf{\textit{K}} & \textbf{\emph{0.02016}}\tabularnewline
\hline 
\end{tabular}
\par\end{centering}
\caption{Pseudo $X^{2}$ values for the original Satterthwaite estimator,
and the three adjustments under examination.}\label{tab: Pseudo X2 values}
\end{table}

While we can confirm that for the three adjustnments compared in this
paper based on the combinations of $K$ and component $\nu$ shown
in the tables above that the adjustment using $C_{opt}$ leads to
the overall smallest $X^{2}$, there is only a small difference when
comparing the results for $C=2.24$, optimized to fit the expected
value for the smallest case $K=2,\nu_{1}=\nu_{2}=1$. It is recommended
to utilize $C=2.24$ for simplicity, but also in order to focus on
the smallest cases, as the asymptotic properties for $K,\overline{\nu}\rightarrow\infty$
take care of any small differences between the estimator and the theoretical
values. That is, it is recommended to use, with 
\[
\overline{\nu}=\frac{\sum_{k=1}^{K}w_{k}\nu_{k}}{\sum_{k=1}^{K}w_{k}}
\]

the adjusted estimator of the effective degrees of freedom given by
\begin{equation}
\nu_{*}\approx\left(1+\frac{2.24}{K\overline{\nu}}\right)^{-1}\frac{\left(\sum_{k=1}^{K}w_{k}S_{k}^{2}\right)^{2}}{\sum_{k=1}^{K}w_{k}^{2}\frac{S_{k}^{4}}{\nu_{k}+2}}
\end{equation}
 for weighted as well as unweighted cases with varying numbers of
components $K$ and for small and \emph{not-so-small} component d.f.
$\nu_{k}.$

\section{Examples}

\subsection{Jackknifing, Balanced Repeated Replications (BRR)}

In the case of single component d.f. $\nu_{k}=1$ and large(-ish)
$K$ representing the (non-trivial) jackknife zones we have no weights
to account for and obtain
\[
\nu_{*}\approx\frac{3}{1+\frac{2.69}{K}}\frac{\left(\sum_{k=1}^{K}\left[T_{\cdot}-T_{JK,k}\right]^{2}\right)^{2}}{\sum_{k=1}^{K}\left[T_{\cdot}-T_{JK,k}\right]^{4}}
\]

since $\nu_{k}+2=3$ for all $k.$ Here, the $T_{JK,k}$ are the pseudo-values
calculated under the jackknife scheme of dropping units, for example
\[
T_{k}=\frac{1}{K}\sum_{j\ne k}f\left(M_{j}\right)
\]
where $M_{j}$ are the means of primary sampling units (clusters/locations/schools),
and $T_{\cdot}=\sum_{k}T_{k}$. In the case of the BRR the $T_{k}$
are the balanced replicated estimates (of the half samples) obtained
by adjusting the weights of pairs of clusters, and recalculating the
statistics $T_{BRR,k}=\sum_{j=1}^{k}w_{kj}f\left(M_{j}\right)$ for
each zone $K$, where the weights $w_{kj}$ are based on a Hadamard
matrix. 

In case of the Jackknife, the Satterthwaite equation, or the improved
approximation presented here, can be directly applied, because the
individual variance components $\left[T_{\cdot}-T_{,JK,k}\right]^{2}$
can be considered independent. In the case of the BRR, there are dependencies
between the $\left[T_{\cdot}-T_{BRR,k}\right]^{2}$ since each of
these contains a linear combination of all cluster (school) means,
since half samples with doubled weights are subtracted from the total
sample mean. In this case, the approach proposed by Kuster (2013)
may be considered, so that the effective d.f. can be computed based
on the $T_{JK,k}$ instead of the $T_{BRR,k}$ since the $BRR$ pseudo
values are (correlated) as these are linear combinations of the independent
$T_{JK,k}.$ 

\subsection{Total Variance under Multiple Imputations}

The total variance of an estimator when imputations is a prime example
of a weighted variance estimate. The total variance as defined by
Rubin (1987) is given by
\[
Var\left(total\right)=Var\left(sampling\right)+\frac{M+1}{M}Var\left(imputation\right)
\]

where the sampling variance may be estimated according to some resampling
scheme (Johnson \& Rust, 1992) and the imputation variance is the
sample standard deviation across $M$ imputation based calculations
of the same statistic.

In this case, the follwing weights are used:

\[
\left(w_{1},w_{2}\right)=\left(1,\frac{M+1}{M}\right)
\]
and equation \ref{eq:new_adjust} can directly be applied.

\subsection{Welch (1947) Test}

The modified Welch (1947) test uses the d.f. $\nu_{1}$ and $\nu_{2}$
of each of the variances of two samples to pool the variance. The
Welch-Satterthwaite formula used in this case can be adjusted using
the improved approximation proposed here. Then we have 
\[
\nu_{*}\approx\left(1+\frac{2.24}{2\overline{\nu}}\right)^{-1}\frac{\left(\frac{1}{N_{1}}S_{1}^{2}+\frac{1}{N_{2}}S_{2}^{2}\right)^{2}}{\frac{S_{1}^{4}}{N_{1}^{2}\left(v_{1}+2\right)}+\frac{S_{1}^{4}}{N_{2}^{2}\left(v_{2}+2\right)}}
\]
where $S_{k}^{2}$ are the sample variances and weights 
\[
w_{k}=\frac{1}{N_{k}}
\]
in this case, for the d.f. $\nu_{*}$ of the sample size weighted
pooled variance used in the Welch test.

\section{Conclusion}

The above developments provide an improved Satterthwaite effective
degrees of freedom approximation for the case of a synthetic variance
estimator that utilizes weighted variance components. a prime example
is the well known formula (Rubin, 1987) for the total variance for
case that imputation are used to estimate sample statistics. The developments
provided in this article generalize and further improve the adjusted
Satterthwaite formula derived by von Davier (2025) for the unweighted
case. It is proposed to estimate the effective degrees of freedom
for the general case of $K\ge2$ and $\nu_{k}\ge1$ using equation
\ref{eq:new_adjust}

\section*{References}

von Davier, M. (2025). An Improved Satterthwaite (1941, 1946) Effective
df Approximation. Journal of Educational and Behavioral Statistics,
0(0). doi:10.3102/10769986241309329

Johnson, E. G., \& Rust, K. F. (1992). Population Inferences and Variance
Estimation for NAEP Data. Journal of Educational Statistics, 17(2),
175--190. doi:10.2307/1165168

Kuster, M. (2013). Applying the Welch-Satterthwaite Formula to Correlated
Errors. NCSLI Measure, 8(1), 42--55. https://doi.org/10.1080/19315775.2013.11721629

Lipsitz, S., Parzen, M., \& Zhao, L. P. (2002). A Degrees-Of-Freedom
approximation in Multiple Imputation. Journal of Statistical Computation
and Simulation, 72(4), 309--318. doi:10.1080/00949650212848

Rubin, D. B., \& Schenker, N. (1986). Multiple Imputation for Interval
Estimation From Simple Random Samples With Ignorable Nonresponse.
Journal of the American Statistical Association, 81(394), 366--374.
doi:10.2307/2289225

Satterthwaite, F. E. (1941). Synthesis of Variance. Psychometrika,
6(5), 309-316. doi:10.1007/BF02288586

Satterthwaite, F. E. (1946). An Approximate Distribution of Estimates
of Variance Components. Biometrics Bulletin, 2(6), 110--114. doi:10.2307/3002019

Welch, B. L. (1947). The generalization of `Student’s’ problem when
several different population variances are involved. Biometrika, 34(1/2),
28-35. doi:10.2307/2332510
\end{document}